\documentclass[epj]{svjour}
\usepackage{amsmath}
\usepackage{graphicx}

\institute{Kazan Federal University, 420008, 18 Kremlyovskaya St., Kazan, Russian Federation\and Institute f\"ur Theoretische Physik III, Ruhr-Universit\"at Bochum, D-44801 Bochum, Germany}
\titlerunning{Dual Features of Magnetic Susceptibility Superconducting Cuprates.}
\authorrunning{M. V. Eremin \and I. M. Shigapov \and I. M. Eremin}

\begin{document}

\author{M. V. Eremin\inst{1} \thanks{%
E-mail address: Mikhail.Eremin@ksu.ru} \and I. M. Shigapov\inst{1} \and I.
M. Eremin\inst{2,1}}
\title{Dual Features of Magnetic Susceptibility in Superconducting Cuprates:
a comparison to inelastic neutron scattering}
\date{}

\abstract{Starting from the generalized $t-J-G$ model Hamiltonian, we analyze
the spin response in the superconducting cuprates taking into account both local
and itinerant spin components which are coupled to each other self-consistently.
We demonstrate that derived expression reproduces the basic observations of neutron
scattering data in $YBa_2Cu_3O_{6+y}$ compounds near the optimal doping level.
\\
\\
(Some figures in this article are in color only in the electronic version)
}

\maketitle

\section{Introduction}

  Theoretical studies of layered cuprates can be broadly separated into two
parts depending on whether the point of consideration is Mott insulator or a
metal. Theories of Mott insulators\cite{theorymott} departs from a Mott
insulator/{Heisen-berg} antiferromagnet (AFM) at half-filling, and addresses
the issue how superconductivity (SC) and metallicity arise upon doping.
Another class of theories \cite{theoryFL} explores the idea that the
system's behaviour is primarily governed by interactions at energies smaller
than the fermionic bandwidth, while contributions from higher energies
account only for the renormalizations of the input parameters. For the
cuprates, the point of departure for such theories is a Fermi liquid (FL) at
large doping, and the issue these theories address is how non-Fermi liquid
physics and unconventional pairing arise upon reducing the doping.
Correspondingly, two approaches are also used for the description of a
dynamic spin susceptibility. In the itinerant case usually one employs the
random phase approximation (RPA) \cite{dahm,Onufrieva,norman,eremin05,chubukov,schnyder} and in a proximity to the
antiferromagnetic state, the spin response above T$_c$ is governed by the
continuum of the antiferromagnetic spin fluctuations (paramagnons). In the
superconducting $d_{x^2-y^2}$-wave state there is a feedback effect of
superconductivity on the spin response which yields a formation of the spin
resonance below T$_c$. While the behaviour of the spin response in the
superconducting state of optimally and overdoped cuprate superconductors can
be qualitatively and even quantitatively understood within this approach,
the normal state data are not entirely captured by the RPA where the
excitations are completely damped and structureless at high energies.%

   Within the localised type of approaches the situation is opposite. In this
case one starts from the undoped situation of a two-dimensional
antiferromagnet and studies how the spin excitations evolve upon introducing
the finite amount of carriers\cite{shimahara,ihle,zavidonov,sherman,prelovsek,yamase,barabanov,vladimirov}.
Here the spin response remains dual in nature as it assumes a mixture of the
local spins described by the superexchange interaction $J$ and the itinerant
carriers with tight-binding energy dispersion. This scenario seems to be
more efficient in describing the normal state spin dynamics, but so far its
application to the superconducting state was rather limited. So called
downward dispersion of neutron scattering intensity is not reproduced in
this approach.%

    Note, the hour-glass-shape dispersion observed in neutron scattering below T$%
_c$ naturally calls for the explanation of the spin response in terms of
dual character of the excitations. While the upward dispersion resembles the
collective spin wave-like branch as in quasi-two dimensional antiferromagnet
with short range spin fluctuation, the downward dispersion in the
superconducting state can be nicely attributed to the feedback effects of
the $d$-wave order parameter on the itinerant component. However, usually an
interaction between both would introduce the repulsion between both branches
and it is not 'a-priori' clear how the total response would look in this
case. In this paper we discuss the possible way how to describe both
components (local and itinerant) on equal footing within one analytical
scheme based on the Green's function method.

\section{Basic equation for quasiparticle operators.}

The starting point for our analysis is the usual $t-J$ type Hamiltonian with
additional density-density interaction term {\small
\begin{equation}
H=\sum_{i,j,\sigma }{t_{ij}\varPsi_{i}^{pd,\sigma }\varPsi_{j}^{\sigma ,pd}}+%
\frac{1}{2}\sum_{i,j}J_{ij}\left[S_{i}S_{j}-\frac{n_{i}n_{j}}{4}\right]+%
\frac{1}{2}\sum_{i,j}G_{ij}\delta _{i}\delta _{j}.
\end{equation}%
} \noindent Here, $\varPsi_{i}^{pd,\sigma}(\varPsi_{j}^{\sigma ,pd})$ are
the creation (annihilation) operators for the composite quasiparticles
within the conduction band of hole-doped cuprates written in terms of
projective operators to obey the no-double occupancy constraint. The second
and third terms describe the superexchange interaction between the spins and
the screened Coulomb interaction of the doped carriers, respectively. $\delta $
is a number of holes per one unit cell. Note that the spin operator
commutes with the density-density interaction and this it will not appear
explicitly in the final expression for the spin susceptibility.

Let us begin with the equation of motion for Fourier-transform of the
quasiparticle operator {\small
\begin{equation}
i\hbar \frac{\partial \varPsi_{k}^{\uparrow ,pd}}{\partial t}=\left[ \varPsi%
_{k}^{\uparrow ,pd},H \right] .
\label{eq:basic}
\end{equation}}
The linearization of the commutator on the rhs of Eq. (\ref{eq:basic}) can
be performed via projection method on the space of creation and annihilation
operators, in a similar way proposed previously \cite{roth,beenen,plakida}.
However, in contrast to the previous approaches\cite{sherman,vladimirov} we
keep also the molecular field terms, which are proportional to the
Fourier-transform of the transverse spin and the density of holes. In
particular, we approximate the commutator in (\ref{eq:basic}) by: {\small
\begin{multline}
i\hbar \frac{\partial \varPsi_{k}^{\uparrow ,pd}}{\partial t}\cong
\varepsilon _{k}\varPsi_{k}^{\uparrow ,pd}+\Delta _{k}\varPsi%
_{-k}^{pd,\downarrow }+\frac{1}{N}\sum_{q}{t^{\prime}_{k-q}\varPsi%
_{k-q}^{\downarrow ,pd}S_{q}^{+}} \\
+\frac{1}{2N}\sum_{q}{t^{\prime}_{k-q}\varPsi%
_{k-q}^{\uparrow ,pd}\delta_{q}}
-\frac{1}{2N}\sum_{q}{J_{q}\varPsi_{k-q}^{\downarrow ,pd}S_{q}^{+}} \\
+\frac{1}{%
N}\sum_{q}{\left[G_{q}-\frac{J_q}{4}\right]\varPsi_{k-q}^{\downarrow ,pd}\delta _{q}},
\label{eq:approximation}
\end{multline}%
} \noindent where $t^{\prime}_{k}=\sum_{j}t_{ij}(1-F_{j})\exp \left(
ikR_{ij}\right) $, $J_{q}=2J_{1}(\cos q_{x}a+\cos q_{y}a)$ and $%
G_{q}=\sum_{j}G_{ij}\exp (iqR_{ij})$ are the Fourier-{trans-forms} of the
hoping integrals, superexchange, and Coulomb interactions, respectively. In
the following we also set the lattice constant $a$ to unity. $N$ is a
number of Cu--sites in the copper--oxygen plane. Collecting terms which are proportional the hopping amplitude, $t_{jl}$, the following expression for the
energy of quasiparticles is obtained 
{\small
\begin{equation}
\varepsilon _{k}=\sum_{l}t_{jl}\left[ \frac{1+\delta }{2}+\frac{2}{1+\delta }%
(1+2F_{l})\left\langle S_{j}^{z}S_{l}^{z}\right\rangle \right] e^{ikR_{jl}}.
\label{eq:ek}
\end{equation}%
} 
The physical meaning of square brackets can be understood as follows. 
The antiferromagnetic spin correlations suppress effective hopping
integral, while ferromagnetic ones increase it.
Thus one should expect that $1+2F_l>0$. Note that overall the effective
nearest neighbour hopping determined as  $\left[ \frac{1+\delta }{2}+\frac{2}{1+\delta }(1+2F_{1})\left\langle S_{0}^{z}S_{1}^{z}\right\rangle \right] $ is analogous to Gutzwiller's
projection $\frac{2\delta }{1+\delta }$ for the hopping integral \cite%
{anderson}. We show below that the exact form of $F_l$ can be rigorously computed via
carriers concentration $\delta$ and the spin-spin
correlations functions, see Eq.(\ref{eq:Ft}).

Note, that the energy dispersion, Eq.(\ref{eq:ek}), can be written in a conventional form
of the tight-binding dispersion, {\small
\begin{equation}
\varepsilon _{k}=2t_{eff}^{(1)}\left( \cos k_{x}a+\cos k_{y}a\right)
+4t_{eff}^{(2)}\cos k_{x}a\cos k_{y}a.  \label{eq:dispersion}
\end{equation}%
} but with effective hopping integrals {\small
\begin{equation}
t_{eff}^{(1)}=t_{1}\left[ P+\frac{1/2+F_{1}}{1+\delta }K_{1}\right] ,
\label{eq:teff1}
\end{equation}%
} {\small
\begin{equation}
t_{eff}^{(2)}=t_{2}\left[ P+\frac{1/2+F_{2}}{1+\delta }K_{2}\right] .
\label{eq:teff2}
\end{equation}%
}
where for shortness we introduce $P=\frac{1+\delta}{2}$, and
the spin-spin correlation functions $K_{n}=4\left\langle S_{0}^{z}S_{n}^{z}\right\rangle$
 are calculated via the spin susceptibility expression in a self-consistent
way for a given doping and temperature.

In the following section we will further proceed with the derivation of the dynamical spin
susceptibility based on the basic equations of motion for the quasiparticle
operators. This approach differs from the memory function method (MFF) %
\cite{sherman,prelovsek,vladimirov} where a linearization of the equations
for the Fourier-transform of the spin operators is not related to the basic
equation for quasiparticale operator (\ref{eq:approximation}). In this
regard our approach is closer to the conventional random phase approximation. However,
it employs projecting operators which fulfil the anticommutator
relation
\begin{equation}
\left\{ \varPsi_{i}^{pd,\uparrow}\varPsi_{j}^{\uparrow ,pd}\right\}
=\left(\frac{1+\delta_i}{2}+S_{i}^{z}\right)\delta _{ij}.
\end{equation}%
which results in richer behavior of the spin susceptibility.
Although this approach requires some intuitive knowledge on the behaviour of
the physical system it has some advantages because (i) both static and dynamic susceptibilities can be calculated within one
approximation scheme, based on (\ref{eq:approximation}), (ii) the extension of the formalism  to
the various symmetry-broken ground states such as superconducting
or spin/charge density wave ordered states is straightforward by adding the corresponding terms in the linearized equation of motion, Eq.(3), and
(iii) the correct asymptotic Fermi-liquid type behaviour of the susceptibilities, when spin-spin
correlation functions become small and the system approaches the conventional Fermi-liquid regime is restored.

In particular, Eq.(3) contains the superconducting gap, $\Delta_{\bf k}$ which mean-field expression can be readily found  {\small
\begin{multline}
\Delta _{k}=\frac{1}{PN}\sum_{k^{\prime }}\biggr[J_{k-k^{\prime
}}\left\langle \varPsi_{k^{\prime }}^{\uparrow ,pd}\varPsi_{-k^{\prime
}}^{\downarrow ,pd}\right\rangle \\
-J_{k-k^{\prime }}\left\langle \varPsi_{k^{\prime }}^{\downarrow ,pd}\varPsi%
_{-k^{\prime }}^{\uparrow ,pd}\right\rangle -G^{\prime}_{k-k^{\prime }}\left\langle %
\varPsi_{k^{\prime }}^{\uparrow ,pd}\varPsi_{-k^{\prime }}^{\downarrow
,pd}\right\rangle \biggm] \\
+\frac{1}{PN}\sum_{k^{\prime }}\biggr[t_{k^{\prime}}\left\langle \varPsi%
_{k^{\prime }}^{\downarrow ,pd}\varPsi_{-k^{\prime }}^{\uparrow
,pd}\right\rangle -t_{k^{\prime }}^{F}\left\langle \varPsi_{k^{\prime
}}^{\uparrow ,pd}\varPsi_{-k^{\prime }}^{\downarrow ,pd}\right\rangle \biggm].
\label{eq:delta}
\end{multline}%
} where $t_{k^{\prime}}=\sum_j{t_{ij}\exp(ikR_{ij})}$,
$t_{k^{\prime}}^F=\sum_j{t_{ij}F_{j}\exp(ikR_{ij})}$ and $G^{\prime}_q=G_q-J_q/4$.
One sees that both superexchange and density-density interactions
are involved. The analysis of this equation for $J_{k-k^{\prime }}>G_{k-k^{\prime }}$, which we assume hereafter, reveals d$_{x^2-y^2}-$wave symmetry to be the most stable solution with $\Delta
_{k}=\Delta (T)(\cos k_{x}a-\cos k_{y}a)/2$.

\section{Dynamic spin susceptibility.}

A hierarchy of constructed equations of motion for determining the spin
response contains five Green's functions. The initial one in the absence of the long
range order ($\langle S_{z}\rangle =0$) has the form {\small
\begin{multline}
\omega \langle \langle S_{q}^{+}|S_{-q}^{-}\rangle \rangle =-\sum_{k^{\prime
}}{(t_{k^{\prime }+q}-t_{k^{\prime }})\langle \langle \varPsi_{k^{\prime
}}^{pd,\downarrow }\varPsi_{k^{\prime }+q}^{\uparrow ,pd}|S_{-q}^{-}\rangle
\rangle }+  \label{eq:GrEq} \\
+\sum_{i,l}{J_{i,l}e^{-iqR_{l}}\langle \langle
S_{l}^{+}S_{i}^{z}-S_{l}^{z}S_{i}^{+}|S_{-q}^{-}\rangle \rangle },
\end{multline}%
} where $t_{k}$ is the usual tight-binding Fourier-transform hopping
integral on the square lattice including the nearest, next-nearest, and
next-next-nearest neighbour hoppings (a bare dispersion). The second term
\begin{equation}
G_{loc}(\omega ,q)=\sum {J_{i,l}e^{-iqR_{l}}\langle \langle
S_{l}^{+}S_{i}^{z}-S_{l}^{z}S_{i}^{+}|S_{-q}^{-}\rangle \rangle }
\end{equation}%
refers to the contribution of the localized spins. Its form is determined by
the same procedure as described in Refs. \cite%
{shimahara,ihle,zavidonov,sherman,barabanov}
{\small
\begin{equation}
\omega G_{loc}(\omega ,q)=-\frac{i}{\pi }J_{1}K_{1}(2-\gamma _{q})+\Omega
_{q}^{2}\langle \langle S_{q}^{+}|S_{-q}^{-}\rangle \rangle
\label{eq:locSp}
\end{equation}%
}
\noindent and $\Omega _{q}$ determines the frequency of collective local
spin fluctuations (magnon-like) {\small
\begin{equation}
\Omega _{q}^{2}=2J_{1}^{2}\alpha (2-\gamma _{q})(\Delta _{sp}-K_{1}(2+\gamma
_{q})),  \label{eq:FoLSF}
\end{equation}%
} \noindent where $\Delta _{sp}$ is a dimensionless parameter of the
so-called spin-gap, \cite{sherman,prelovsek} , $\gamma _{q}=\cos q_{x}a+\cos
q_{y}a$, and $\alpha $ is a decoupling parameter, controlled by the sum rule
$\langle S_{i}^{+}S_{i}^{-}\rangle =\frac{1}{2}(1-\delta )$, which is
about $1.4$.

The first term on the right hand side of Eq. (\ref{eq:GrEq}) \linebreak
{\small
\begin{equation*}
G_{it}(\omega ,q)=\frac{1}{2}\sum_{k^{\prime }}{(t_{k^{\prime
}+q}-t_{k^{\prime }})\langle \langle \varPsi_{k^{\prime }+q}^{\uparrow ,pd}%
\varPsi_{k^{\prime }}^{pd,\downarrow }-\varPsi_{k^{\prime }}^{pd,\downarrow }%
\varPsi_{k^{\prime }+q}^{\uparrow ,pd}|S_{-q}^{-}\rangle \rangle }
\end{equation*}}
is determined by the dynamics of itinerant spins. To compute this
function we use the following exact relations\cite%
{eremin1,eremin2}
{\small
\begin{equation}
\sum_{k}{\varPsi_{k}^{pd,\downarrow }\varPsi_{k+q}^{\uparrow ,pd}} = 0,
\label{eq:ExactRelat1}
\end{equation}
\begin{equation}
\sum_{k}{\varPsi_{k+q}^{\uparrow ,pd}\varPsi_{k}^{pd,\downarrow }} =%
S_{q}^{+}.
\label{eq:ExactRelat2}
\end{equation}}
Using Eq. (\ref{eq:approximation}) we construct further equations for the Green's functions
{\small
\begin{equation}
\langle \langle \varPsi_{k}^{pd,\downarrow} \varPsi_{k+q}^{\uparrow ,pd} | S_{-q}^{-}\rangle\rangle
 = \frac{i}{2\pi} \chi_{k,q} + \frac{1}{N}\eta^{\prime}_{k,q} \langle \langle S_{q}^{+}|S_{-q}^{-}\rangle \rangle
 + \frac{\zeta_{k,q}}{N} D^{\prime}\left( \omega,q\right),
\label{eq:Eq1}
\end{equation}
\begin{equation}
\langle \langle \varPsi_{k+q}^{\uparrow ,pd} \varPsi_{k}^{pd,\downarrow} | S_{-q}^{-} \rangle \rangle
= -\frac{i}{2\pi} \chi_{k,q} - \frac{1}{N}\eta^{\prime \prime}_{k,q}\langle \langle S_{q}^{+}|S_{-q}^{-}\rangle \rangle
 - \frac{\zeta_{k,q}}{N} D^{\prime \prime}\left( \omega,q\right).
\label{eq:Eq2}
\end{equation}%
}
Observe that here the new Green's functions appear
{\small
\begin{eqnarray}
D^{\prime}\left( \omega
,q\right) =-\sum_{k^{\prime }}{(\varepsilon _{k^{\prime }+q}-\varepsilon
_{k^{\prime }})\langle \langle \varPsi_{k^{\prime }}^{pd,\downarrow }\varPsi%
_{k^{\prime }+q}^{\uparrow ,pd}|S_{-q}^{-}\rangle \rangle }, \\
D^{^{\prime \prime }}\left( \omega ,q\right) =\sum_{k^{\prime }}{%
(\varepsilon _{k^{\prime }+q}-\varepsilon _{k^{\prime }})\langle \langle %
\varPsi_{k^{\prime }+q}^{\uparrow ,pd}\varPsi_{k^{\prime }}^{pd,\downarrow
}|S_{-q}^{-}\rangle \rangle }.
\end{eqnarray}}
Combining Eqs.(\ref{eq:ExactRelat1})-(\ref{eq:Eq2}) we find
{\small
\begin{equation}
D^{\prime }(\omega ,q)=-\left\{ \frac{iN}{2\pi }\chi (\omega ,q)+\eta
^{\prime }(\omega ,q)\langle \langle S_{q}^{+}|S_{-q}^{-}\rangle \rangle
\right\} /\zeta (\omega ,q)
\label{eq:D1}
\end{equation}%
}
and
{\small
\begin{equation}
D^{\prime \prime }(\omega ,q)=-\left\{ \frac{iN}{2\pi }\chi (\omega ,q)+%
\left[ 1+\eta ^{^{\prime \prime }}(\omega ,q)\right] \langle \langle
S_{q}^{+}|S_{-q}^{-}\rangle \rangle \right\} /\zeta (\omega ,q),
\label{eq:D2}
\end{equation}
}
where for $T>T_c$
{\small
\begin{equation}
\chi (\omega ,q)=\sum_{k}{\chi_{k,q}}
=\frac{1}{N}\sum \frac{n_{k+q}-n_{k}}{\omega +\varepsilon
_{k}-\varepsilon _{k+q}},
\end{equation}%
}
{\small
\begin{equation}
\zeta (\omega ,q)=\sum_{k}{\zeta_{k,q}}
=\frac{1}{N}\sum \frac{1}{\omega +\varepsilon
_{k}-\varepsilon _{k+q}},
\end{equation}
}
{\small
\begin{eqnarray}
\eta ^{\prime }(\omega ,q)=\sum_k{\eta^{\prime}_{k,q}}
&=&\frac{1}{2}J_{q}\chi (\omega ,q) \nonumber \\
&-&\frac{1}{N}\sum \frac{t^{\prime}_{k+q}n_{k+q}-t^{\prime}_{k}n_{k}}{\omega +\varepsilon
_{k}-\varepsilon _{k+q}},
\end{eqnarray}%
}
{\small
\begin{eqnarray}
\eta ^{^{\prime \prime }}(\omega ,q)=\sum_{k}{\eta^{\prime \prime}_{k,q}}
&=&\eta ^{\prime }(\omega ,q) \nonumber \\
&+& \frac{1}{N}%
\sum \frac{ P (t^{\prime}_{k+q}-t^{\prime}_{k})-\omega}{\omega +\varepsilon
_{k}-\varepsilon _{k+q}}.
\end{eqnarray}%
}
We remind that the difference between $t_{k}$ and $t^{\prime}_{k}$ is expressed below Eq.(3) and $n_k = P f_k$ where $f_k$ is a Fermi function.
Taking into account that \noindent $\sum_{k^{\prime }} {(\varepsilon _{k^{\prime
}+q}-\varepsilon _{k^{\prime }})=0}$ it is easy to prove that $D^{\prime
}\left( \omega ,q\right) =D^{^{\prime \prime }}\left( \omega ,q\right)$.
This means that Eqs.(\ref{eq:D1}) and (\ref{eq:D2}) are consistent with each other if the following
relation holds
{\small
\begin{equation}
 1+\frac{1}{N} \sum_{k} \frac{ P (t^{\prime}_{k+q}-t^{\prime}_{k})-\omega}{\omega
+ \varepsilon _{k}-\varepsilon_{k+q}}=0.
\end{equation}
} %
This property can be used to express the projection parameter $F_{i}$ via the
spin-spin correlation function $K_{i}$. For this we add and subtract
$\varepsilon_k-\varepsilon_{k+q}$ in the numerator of Eq.(26) which yields the identity
{\small
\begin{equation}
\sum_{k}\frac{ P (t^{\prime}_{k+q}-t^{\prime}_{k})-\varepsilon _{k+q}+\varepsilon_{k}}{\omega
+ \varepsilon _{k} - \varepsilon_{k+q}}=0.
\end{equation}
} 
Note that the remaining sum vanishes at
any frequency if $P(t^{\prime}_{k+q}-t^{\prime}_{k})-\varepsilon_{k+q}
+\varepsilon_k=0$. Equating factors in front of the independent trigonometric cosine functions results in the relation between $F_i$ and corresponding spin-spin correlation functions
{\small
\begin{equation}
  F_{i} = \frac{-K_i}{(1+\delta)^2+2K_i}
  \label{eq:Ft}
\end{equation}
}
which holds for any value of $i$. Recalling that $1+2F_1>0$, we find that
$(1 + \delta)^2>2|K_1|$. This condition implies that  the concentration of carriers should
be larger than some critical value $\delta_0$ to have the metallic behaviour of the system.   

Having $D^{\prime}\left(\omega, q \right)$, $D^{\prime \prime}\left(\omega, q \right)$
and using Eqs.(\ref{eq:Eq1}), (\ref{eq:Eq2}) it is straightforward to
find $G_{it}(\omega ,q)$ and $\langle \langle S_{q}^{+}|S_{-q}^{-}\rangle
\rangle $. The obtained system of coupled equations
allow us to get a closed form for the Green's function $\langle \langle
S_{q}^{+}|S_{-q}^{-}\rangle \rangle $ {\it i.e.}, for the spin
susceptibility
{\small
\begin{equation}
\chi _{total}^{+,-}=\frac{\omega \chi (\omega ,q)\zeta _{t}(\omega
,q)-[\omega \chi _{t}(\omega ,q)+2J_{1}K_{1}(2-\gamma _{q})]\zeta (\omega ,q)%
}{\omega \left[ 1/2+\eta (\omega ,q)\right] \zeta _{t}(\omega ,q)+[\Omega
_{q}^{2}-\omega ^{2}-\omega \eta _{t}(\omega ,q)]\zeta (\omega ,q)}.
\label{eq:chiTotal}
\end{equation}}%
Here,
{\small
\begin{multline}
\eta (\omega ,q)=\frac{1}{2}J_{q}\chi (\omega ,q) \\
-\frac{1}{N}\sum_{k}{ \frac{t_{k+q}^{\prime}(n_{k+q}-P/2)-t_{k}^{\prime }(n_{k}-P/2)}
{\omega +\varepsilon_{k}-\varepsilon_{k+q}}},
\end{multline}}
and
{\small
\begin{eqnarray}
\zeta _{t}(\omega ,q) &=&\frac{P}{N}\sum_{k}(t_{k+q}-t_{k})\zeta_{k,q},
\notag \\
\chi _{t}(\omega ,q) &=&\frac{P}{N}\sum_{k}(t_{k+q}-t_{k})\chi_{k,q},
\label{eq:Eq_t}
\\
\eta _{t}(\omega ,q) &=&\frac{P}{N}\sum_{k}(t_{k+q}-t_{k})\eta_{k,q}.
\notag
\end{eqnarray}%
}
Note that the analytic continuation $\omega =\omega +i0^+$ used in the derivation is replaced
in the numerical calculations by $\omega=\omega+i\varGamma$ where $\varGamma$ is a small numerical factor. It is frequency independent and
resembles the effect of the non-magnetic impurity
scattering. \cite{norman,eremin05,chubukov,schnyder}

Let us turn now to the situation of the superconducting state, $T<T_c$. Performing the
Bogolyubov transformations for the quasiparticle states and obtaining the
new Eigenenergies $E_{k}$, we then employ the new projecting operators, $\tilde{%
\Psi}$ that takes into account the $\left\{ u,v\right\} $ Bogolyubov
coefficients. This procedure is standard in the theory of superconductivity
and thus we skip the details. Note that overall expression for the spin
response retains its form except that the entering functions such as $\chi (\omega ,q)$ modify:
{\small
\begin{multline}
\chi (\omega ,q)=\frac{1}{N}\sum_{k}{\chi _{kq}}=\frac{P}{N}\sum {S_{xx}%
\frac{f_{k+q}-f_{k}}{\omega +E_{k}-E_{k+q}}}+  \label{eq:BCS} \\
+\frac{P}{N}\sum {S_{yy}\frac{f_{k}-f_{k+Q}}{\omega -E_{k}+E_{k+q}}}+ \\
+\frac{P}{N}\sum {S_{yx}^{(-)}\frac{f_{k+q}+f_{k}-1}{\omega -E_{k}-E_{k+q}}}+
\\
+\frac{P}{N}\sum {S_{xy}^{(+)}\frac{1-f_{k+q}+f_{k}}{\omega +E_{k}+E_{k+q}},}
\end{multline}}%
\noindent which resembles BCS-like expression for the spin susceptibility for non-interaction electrons and the
Fermi function $f_k$ contains now the new eigenenergies. Note that the
superconducting gap is determined by the numerical solution of Eq.(9).
For the sake of simplicity, we further use the following abbreviations
for the Bogolyubov coherence factors:
{\small
\begin{eqnarray}
S_{xx}=x_{k}x_{k+q}+z_{k}z_{k+q},\quad S_{yy} &=&y_{k}y_{k+q}+z_{k}z_{k+q},
\notag \\
S_{xy}^{(+)}=x_{k}y_{k+q}-z_{k}z_{k+q},\quad S_{yx}^{(-)}
&=&y_{k}x_{k+q}-z_{k}z_{k+q},
\end{eqnarray}}%
where
{\small
\begin{eqnarray}
x_{k} &=&\frac{1}{2}\left[ 1+\frac{\varepsilon _{k}-\mu }{E_{k}}\right]
,\quad y_{k}=\frac{1}{2}\left[ 1-\frac{\varepsilon _{k}-\mu }{E_{k}}\right] ,
\notag \\
z_{k} &=&\frac{\Delta _{k}}{2E_{k}},\quad E_{k}=\sqrt{(\varepsilon _{k}-\mu
)^{2}+\lvert \Delta _{k}\rvert ^{2}}.
\end{eqnarray}}%
\noindent Then the function $\eta (\omega ,q)$ gets the form
{\small
\begin{multline}
\eta (\omega ,q)=\frac{1}{2}J_{q}\chi (\omega ,q)-\frac{P}{N}\biggr \{
\label{eq:eta`} \\
\phantom{+}\sum_{k}{\ S_{xx}\frac{t_{k+q}^{\prime
}(f_{k+q}-1/2)-t_{k}^{\prime }(f_{k}-1/2)}{\omega +E_{k}-E_{k+q}}}+ \\
+\sum_{k}{\ S_{yy}\frac{t_{k+q}^{\prime }(1/2-f_{k+q})-t_{k}^{\prime
}(1/2-f_{k})}{\omega -E_{k}+E_{k+q}}}+ \\
+\sum_{k}{\ S_{yx}^{(-)}\frac{t_{k+q}^{\prime }(f_{k+q}-1/2)-t_{k}^{\prime
}(1/2-f_{k})}{\omega -E_{k}-E_{k+q}}}+ \\
+\sum_{k}{\ S_{xy}^{(+)}\frac{t_{k+q}^{\prime }(1/2-f_{k+q})-t_{k}^{\prime
}(f_{k}-1/2)}{\omega +E_{k}+E_{k+q}}}\biggm \}
\end{multline}}%
\noindent where $P$ is again the thermodynamic average of anticommutator \linebreak $%
\langle \varPsi_{i}^{pd,\sigma }\varPsi_{i}^{\sigma ,pd}+\varPsi_{i}^{\sigma
,pd}\varPsi_{i}^{pd,\sigma }\rangle =\frac{1+\delta }{2}$ which  is the same as in the normal state.
As mentioned above the anticommutation relations for projecting
operators are not simple fermionic ones. As a result the last terms which appear in Eq. (\ref%
{eq:eta`}) correspond to an effective molecular field of kinematic origin
which is a result of the projection origin of the operators (strong correlation effects).

 The function $\zeta (\omega ,q)$ is written as follows:
{\small
\begin{multline}
\zeta (\omega ,q)=\frac{1}{N}\sum_{k}{\zeta _{kq}}=\frac{1}{N}\sum_{k}{\
\frac{S_{xx}}{\omega +E_{k}-E_{k+q}}}+  \label{eq:zeta} \\
+\frac{1}{N}\sum_{k}{\ \frac{S_{yy}}{\omega -E_{k}+E_{k+q}}}+ \\
+\frac{1}{N}\sum_{k}{\ \frac{S_{yx}^{(-)}}{\omega -E_{k}-E_{k+q}}}+ \\
+\frac{1}{N}\sum_{k}{\ \frac{S_{xy}^{(+)}}{\omega +E_{k}+E_{k+q}}}.
\end{multline}}%
\noindent At the same time the terms in Eq.(\ref{eq:chiTotal} which describe the coupling between the
magnetizations of the itinerant and localized spins remains the same as in the normal state, see
(\ref{eq:Eq_t}). Observe also that Eqs. (\ref{eq:BCS}), (\ref{eq:eta`}),
(\ref{eq:zeta}) refer to itinerant spin-component of the spin susceptibility. If
we assume for the moment that no hopping terms ( i.e.$t_{1}=t_{2}=0$) of
conduction electrons (holes) are present, Eq. (\ref{eq:chiTotal})
reduces to
{\small
\begin{equation}
\chi _{local}^{+,-}=\frac{2J_{1}K_{1}(2-\gamma _{q})}{\omega ^{2}-\Omega
_{q}^{2}}.
\end{equation}}%
This expression is identical to those found previously and is widely used to
describe the lightly doped cuprates\cite{ihle,zavidonov,sherman,prelovsek,barabanov}. It is remarkable that
magnetism of localized spins in Eq. (\ref{eq:chiTotal}) is strongly suppressed (or in other words "frozen out")
due to the superconducting gap, which is
naturally incorporated in expression for $\zeta (\omega ,q)$. In the
opposite limit, when the spin-spin correlation function is small and
conduction electrons bandwidth is large enough ($\zeta (\omega ,q)$ is
small), Eq. (\ref{eq:chiTotal}) becomes similar (but not identical) to the one obtained
previously  in the so-called generalized random phase
approximation (GRPA) scheme for Hubbard model in the normal state \cite{auslender}.
In contrast to the previously obtained expression, see Ref. \cite{auslender},
Eq.(29) obeys  the electron-hole symmetry  which is especially important for the superconducting state.

In addition note that Eq.(\ref{eq:chiTotal}) contains contributions from both the itinerant and the local
components of the spin susceptibility, which are mutually coupled in a non-trivial way. In fact, these components
cannot be easily separated and should be treated as collective spin response of the entire system. The energy
position of the spin excitations is obtained by analysing the pole structure
of the denominator of Eq.(\ref{eq:chiTotal}). In quasi-localised spins regime
this equation corresponds to a short-range magnon-type oscillations with an
upward dispersion near the antiferromagnetic wave vector $Q=(\pi ,\pi )$. In
the case of the itinerant spins the expression (\ref{eq:chiTotal}) yields a
Stoner continuum with the overdamped paramagnon excitation.

\section{Numerical results and comparison to experimental data}

In the following we present the numerical results using Eq. (\ref%
{eq:chiTotal}) for the normal and superconducting state. As we pointed out above the superconducting gap equations yields
$d_{x^{2}-y^{2}}$-wave symmetry of the superconducting gap for $J_{k-k^\prime}>G_{k-k^\prime}$. Therefore we approximate the superconducting gap in the form $\Delta
_{k}=\Delta _{0}\left( \cos k_{x}a-\cos k_{y}a\right) /2$. The magnitude of
$\Delta _{0}\approx 30meV$ is obtained from the temperature
dependencies of the nuclear relaxation rate \cite{mayer} and superfluid
density \cite{eremin3} for $YBa_{2}Cu_{3}O_{7}$. The energy dispersion is
given by (\ref{eq:dispersion}) and we employ the following minimal set of
effective hoping parameters (in meV): $t_{eff}^{(1)}=250$, $%
t_{eff}^{(2)}=-50 $ which reproduce the observed Fermi surface for optimally-doped cuprates near the
optimal doping level.
\begin{figure*}[th]
\noindent\centering{\ \includegraphics[width=85mm]{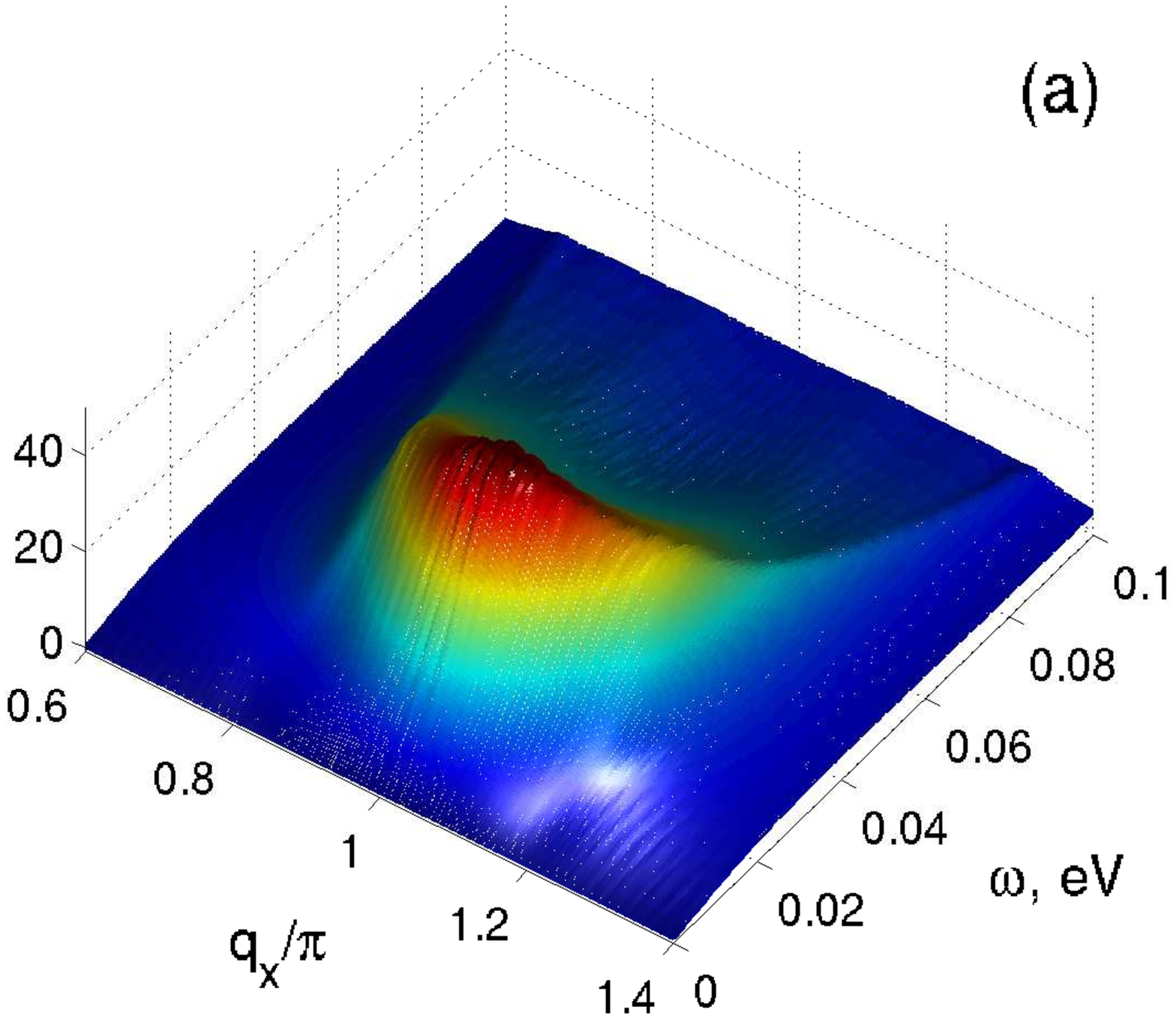} %
\includegraphics[width=85mm]{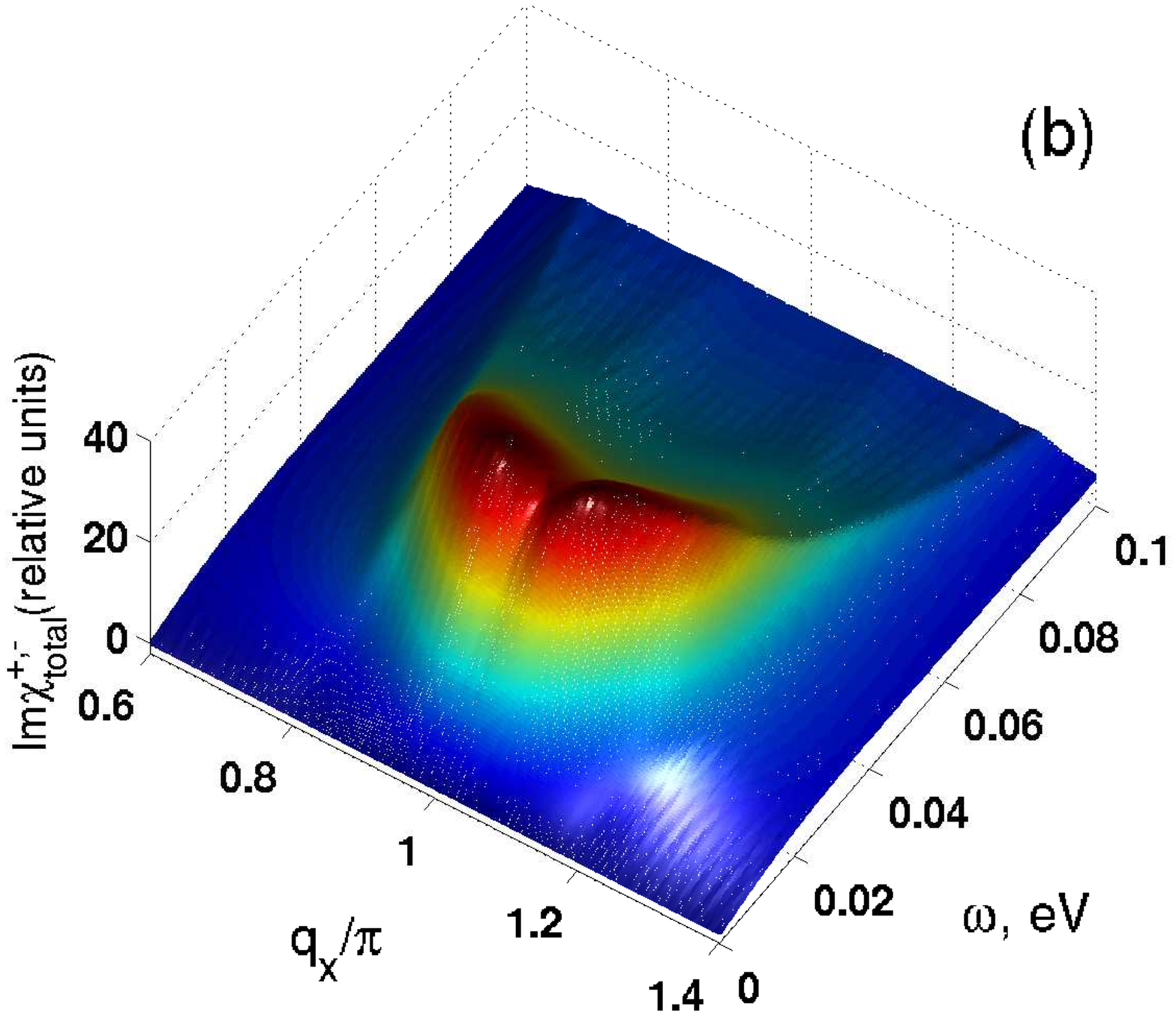} }
\caption{Calculated imaginary part of the spin susceptibility for the normal
state (T=100K) near the wave vector $Q=(\protect\pi ,\protect\pi )$ as a
function of $\protect\omega $ (in eV) and $q_{x}$ [$q_{y}=\protect\pi $] for
$\varGamma=6$meV \textbf{(a)} and that using Eq.(\ref{eq:Gamma}) with $\varGamma_{1}=6meV$, $\varGamma_{2}=6meV$, $\protect%
\xi =0.05\protect\pi $. \textbf{(b)}}.
\label{fig:Normal}
\end{figure*}
\begin{figure*}[tbh]
\noindent\centering{\ \includegraphics[width=120mm]{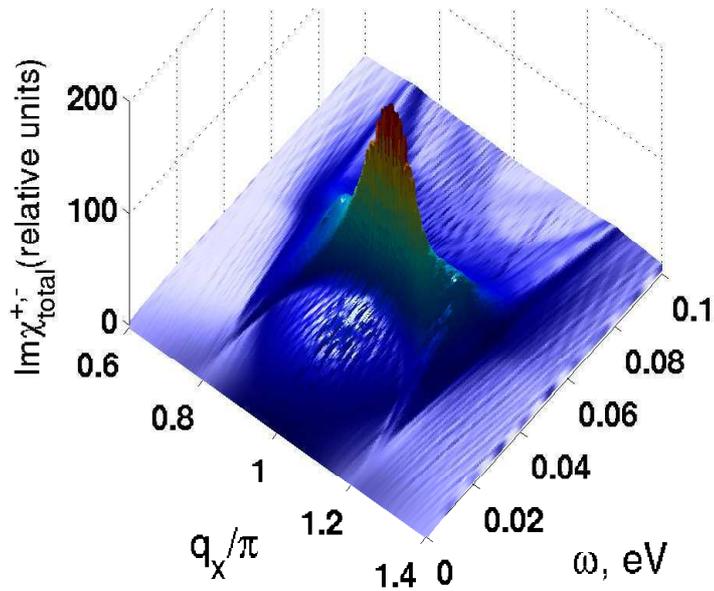} %
}
\caption{Calculated imaginary part of spin susceptibility for the
superconducting state (T=10K) near the wave vector $Q=(\protect\pi ,\protect%
\pi )$ as a function of $\protect\omega $ (in eV) and $q_{x}$ for $q_{y}=%
\protect\pi $.
}
\label{fig:Superconduct}
\end{figure*}

The calculated imaginary part of susceptibility in the
normal phase for $T=100K$ is shown in Fig. \ref{fig:Normal}. The chosen
parameters are: $J_{1}=100meV$, $\Delta _{sp}=0.18$, and $\varGamma=6meV$.
Values $F_{1}=0.15$ and $F_{2}=-0.02$ was estimated using Eq (\ref{eq:Ft}) at $K_{1}=-0.2$,
$K_{2}=0.04$ and $\delta=0.3$.
Observe that the imaginary part of total susceptibility has a visible upward
dispersion, which can be attributed to a magnon-like mode damped at high
frequencies due to coupling to the itinerant carriers. Note that the visible
dispersion of the spin excitations centered at the antiferromagnetic wave
vector \textbf{Q}, is in contrast to that found within a simple RPA
expression\cite{reznik}. There the spin excitations are almost structureless
in the normal state and refer to the overdamped Stoner continuum. Another
important feature is that the spin excitations remain commensurate despite
the significant hole doping as shown in Fig.\ref{fig:Normal}(a). This is due
to the fact that the spin excitations has a character of the almost
localised magnetic modes which are less sensitive to the degree of the Fermi
surface nesting. One has, however, to keep in mind that in our calculations
we take $\varGamma=6meV$ as a constant. In principle in a real physical system this value  may become momentum and frequency
dependent. In particular, assuming the coupling of the
quasiparticles to the spin excitations, the quaiparticle lifetime will
become anisotropic as a function of the Fermi surface angle. Quasiparticles
connected by the antiferromagnetic momentum will scatter stronger as
compared to those located near the diagonal of the BZ. As a result, the
imaginary part of the self-energy (which affects $\varGamma$) will be larger around the wave vector
\textbf{Q} and is smaller away from it. To demonstrate this effect
we approximate $\varGamma$ by a Lorenzian in the form
\begin{equation}
\varGamma(q)=\varGamma_{1}+\varGamma_{2}\frac{(\xi )^{2}}{(|q_{x}|-\pi
)^{2}+(|q_{y}|-\pi )^{2}+(\xi )^{2}}.  \label{eq:Gamma}
\end{equation}%
\noindent where $\xi$ is a magnetic correlation length. As one sees the effect of \textbf{q}-dependence of the damping $%
\varGamma $ yields a dip in imaginary part of susceptibility at $(\pi ,\pi )$
as shown in Fig.\ref{fig:Normal}(b) and to the incommensurate magnetic
fluctuations.

In Fig. \ref{fig:Superconduct} we show the imaginary part of the
spin susceptibility in the superconducting state with $d_{x^2-y^2}$-wave
symmetry. Observer the strong renormalization of the entire spectrum of the
spin excitations. As a consequence of the $d$-wave symmetry of the order
parameter there is a resonance mode forming at energies well below $|\Delta_{%
\mathbf{k}} + \Delta_{\mathbf{k+Q}}| \sim 60$meV. Away from the
antiferromagnetic momentum, the excitations shows the $X-$shape structure
which appears only in the superconducting state. Due to the delicate balance
between the spin-gap parameter $\Delta _{s}$ and the twice superconducting
gap, $2\Delta _{0}$, the frequency position of the resonance peak in
superconducting state is almost unchanged with regard to the normal state
value. However, its position is still below the onset of particle-hole
continuum which starts approximately at $2\Delta_0$. Taking the constant
energy cuts of the excitations shown in Figs. \ref{fig:Normal} and \ref%
{fig:Superconduct} one finds that the results are very similar to those
observed in inelastic neutron scattering in $YBa_{2}Cu_{3}O_{6.92}$ and $%
YBa_{2}Cu_{3}O_{6.97}$ (see Fig. 8) in P. Bourges Rev. \cite{bourge}. Note
that the calculated upward and downward dispersions shown in Fig. \ref%
{fig:Superconduct} look similar to those found experimentally in $%
YBa_{2}Cu_{3}O_{6.95}$ (see Fig. 3(c) in \cite{reznik}), (Fig. 5(a)-(b) in
Ref. \cite{bourge}), $YBa_{2}Cu_{3}O_{6.85}$ (Fig. 4(a) in Ref. \cite%
{pailhes}), $YBa_{2}Cu_{3}O_{6.6}$ (Fig.5 in Ref. \cite{arai}), and $%
YBa_{2}Cu_{3}O_{6.5}$ (Fig. 5 in Ref.\cite{stock}). The coexistence of both
downward and upward branches can be clearly seen in Fig. \ref%
{fig:Superconduct}. While the downward dispersion refers to the itinerant
component of the spin response and the evolution of $d$-wave gap on the
Fermi surface, the upward dispersion originates mostly from magnon-like
short-range antiferromagnetic fluctuations. Important to note here is that
these dual features are obtained using the general expression for the spin
susceptibility which accounts for both local and itinerant spin excitations.
\begin{figure}[t]
\noindent\centering{%
\includegraphics[width=80mm,height=70mm]{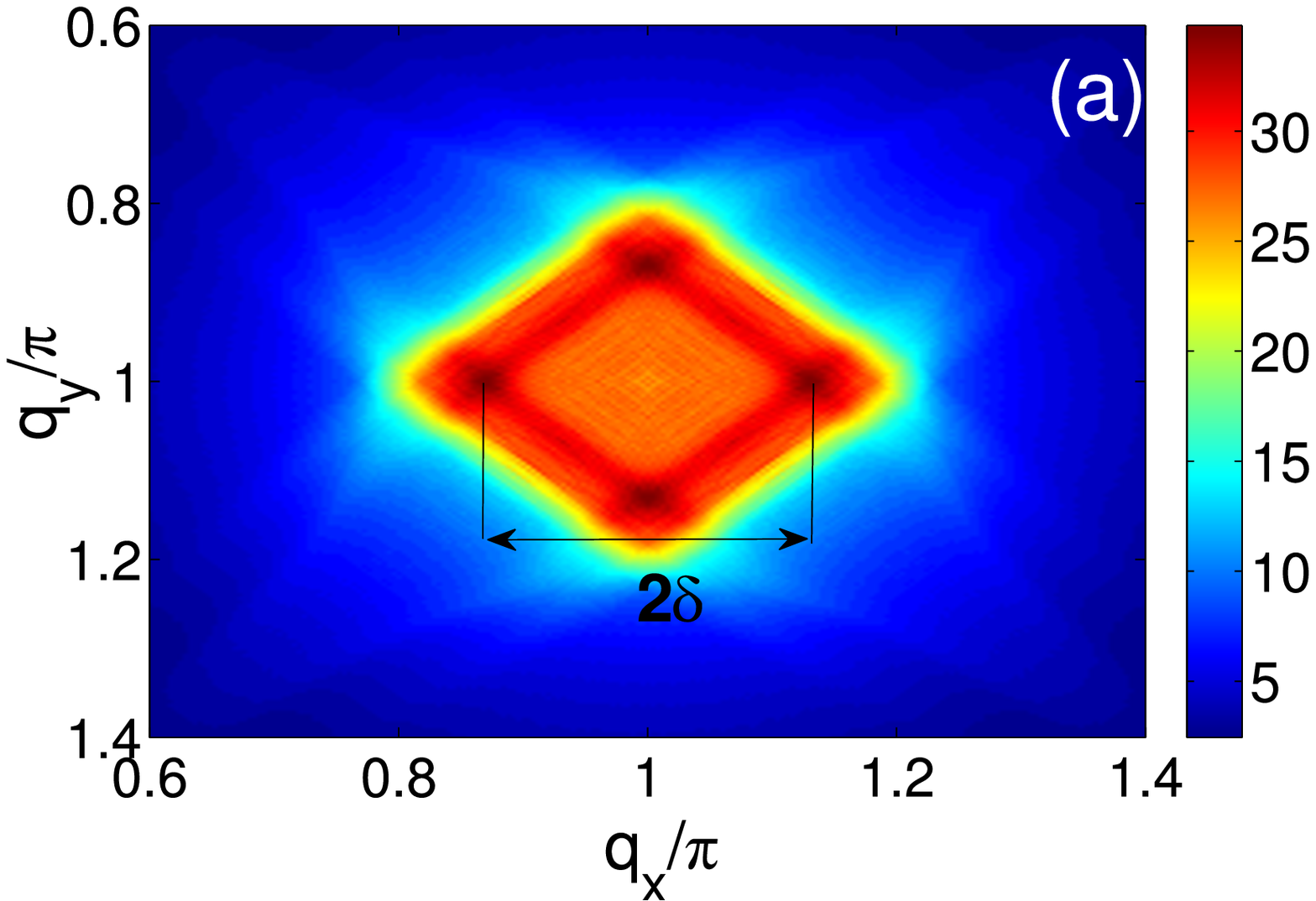} %
\includegraphics[width=80mm,height=70mm]{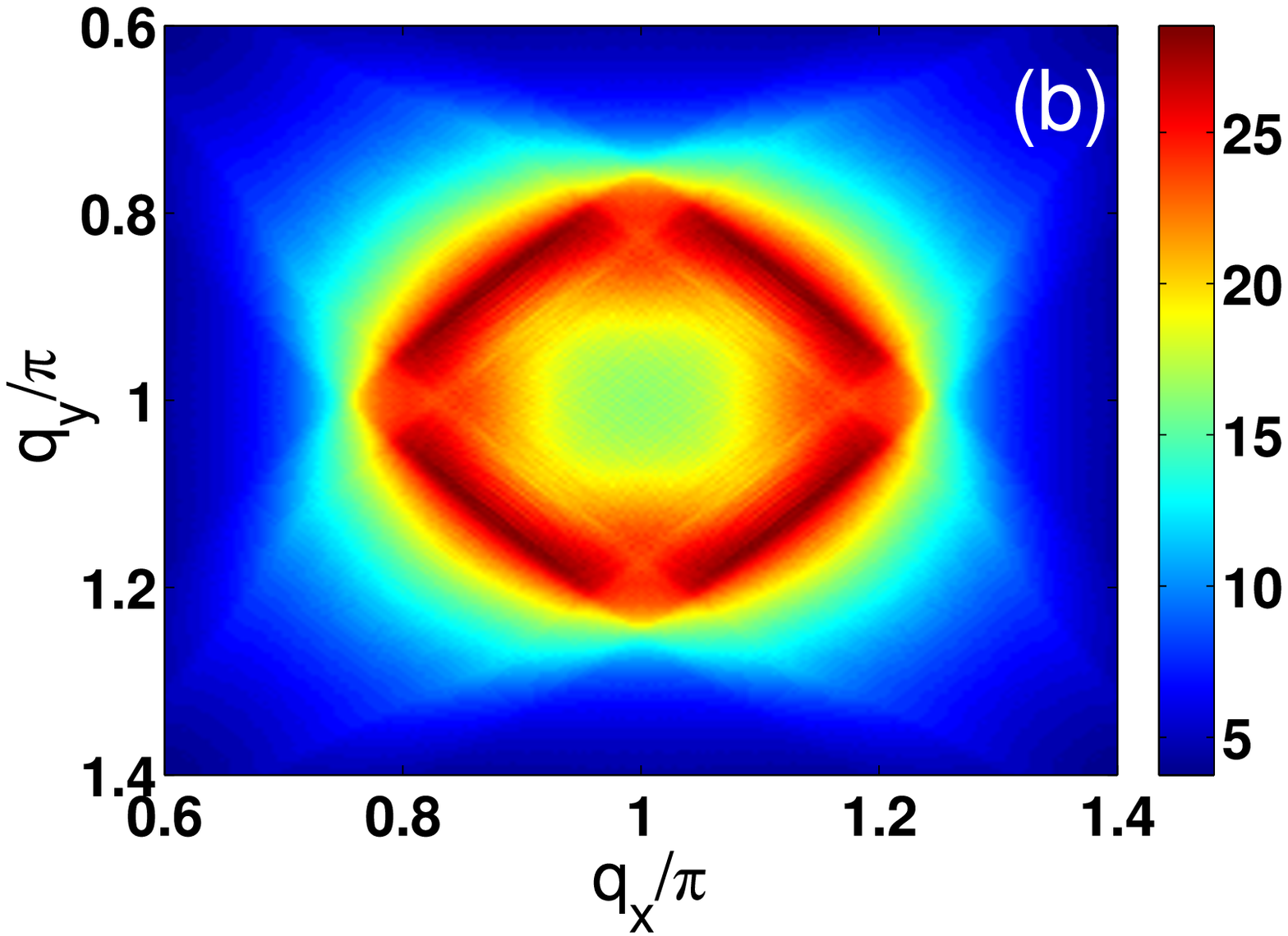} }
\caption{Intensity plot of the imaginary part of the susceptibility $Im
\protect\chi ^{+,-}(\protect\omega ,q)$ around wave vector $Q=(\protect\pi ,
\protect\pi )$ for $\hbar \protect\omega =40meV$ \textbf{(a)} and $\hbar
\protect\omega =60meV$ \textbf{(b)},$\varGamma=6meV$.}
\label{fig:IntensityPlot}
\end{figure}

\begin{figure}[th]
\noindent \includegraphics[width=80mm]{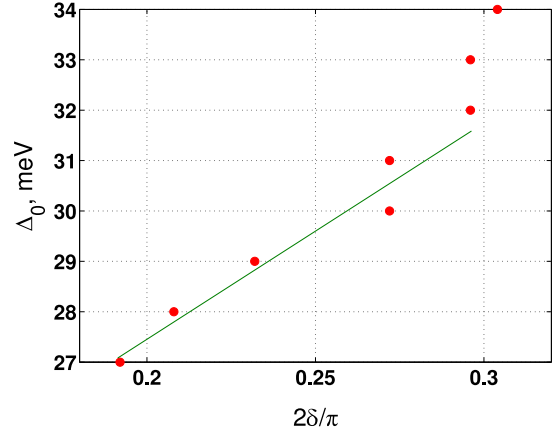}
\caption{The incommensurability parameter, $2\protect\delta/\protect\pi $
plotted as a function of the superconducting gap amplitude $\Delta _{0}$.
The red points refer to the calculated values using Eq. (\ref{eq:chiTotal})
and the solid line is a guide to the eye.}
\label{fig:Incommensurate}
\end{figure}

In Fig.\ref{fig:IntensityPlot} we show the intensity plot of the
imaginary part of the susceptibility $Im\chi ^{+,-}(\omega ,q)$ as a
function of \textbf{q} around the wave vector $Q=(\pi,\pi )$ in the
superconducting state for the energies above and below the resonance energy
positions, $\omega_{res}$. Below $\omega_{res}$ in agreement with
experimental observation \cite{reznik} and RPA calculations \cite{eremin05}
we have found quadratic in \textbf{q} downward dispersion with dominant
peaks along the bond directions. The latter is due to fact that there is
more phase space available for $(\pi,\pi+\delta)$ directions as they are
closer to the resonance at $(\pi,\pi)$ than those at the diagonal wave
vector at this energy. At the same time, the upward branch above $%
\omega_{res}$ has almost circular symmetry around $Q_{AF}$ which is the same
as in the normal state and agrees with experimental observation \cite{reznik}%
. In order to see the evolution of the incommensurability with the
superconducting gap we plot $\Delta_0$ vs $2\delta/\pi$ in Fig. \ref{fig:Incommensurate}.
Observe that for increasing $\Delta_0$ the incommensurability parameter, $%
\delta$, computed for $\omega=40meV$ increases. This is an indication that
the incommensurability for $\omega < \omega_{res}$ refers to the spin
exciton dispersion which is connected to the magnitude of the
superconducting gap. Namely, the increase of $\Delta_0$ shifts the position
of the resonance peak to higher energies. For a given $\omega<\omega_{res}$
this increase implies that the itinerant spin excitations arises from the
scattering of the quasiparticle states located at the Fermi surface points
which are closer and closer to the diagonals of the BZ, \textit{i.e.}
smaller \textbf{q}$_i$ (larger $2\delta$). Note that this observation is in
direct correspondence with the result found previously in experimental data
(see Fig. 25 in Ref. \cite{dai}). At certain point, when the resonance
energy is large enough, the only electron-hole scattering available
originates from the nodal points where the superconducting gap is zero. At
this moment, the incommensurability starts to be independent on the $%
\Delta_0 $, the effect clearly visible in Fig.\ref{fig:Incommensurate}. This
observation confirms the itinerant origin of the lower branch of the
dispersion of the spin excitations. Note that the incommensurability of the
spin excitations for $\omega>\omega_{res}$ is not sensitive to the magnitude
of the superconducting gap as it arises from the localised magnetic
excitations. In this case it remains a constant (not shown) independent on
the size of the superconducting gap.
\begin{figure}[h]
\noindent \includegraphics[width=80mm]{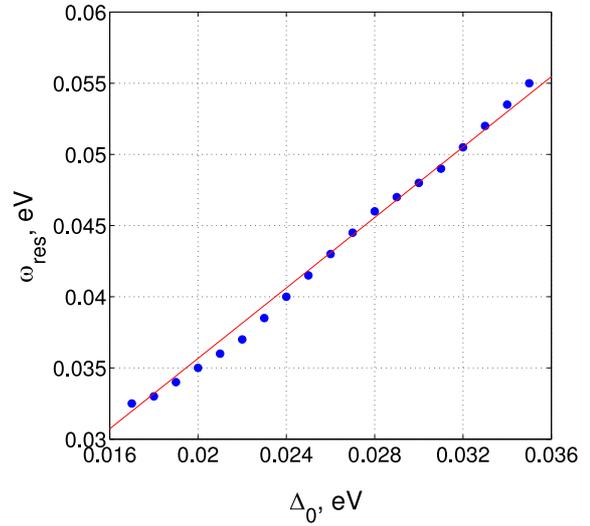}
\caption{Calculated relationship between the resonance energy (in meV) and
superconductivity gap $\Delta _{0}$.}
\label{fig:GapEnergy}
\end{figure}

In Fig.\ref{fig:GapEnergy} we show the relation between the
resonance energy and superconductivity gap $\Delta _{0}$. Despite the fact
that this observation is a direct consequence of the resonance being the
spin exciton we show it here as it agrees fairly well with experimental data
(see, for example, Fig. 26 in Ref. \cite{dai}).
\begin{figure}[h]
\includegraphics[width=80mm]{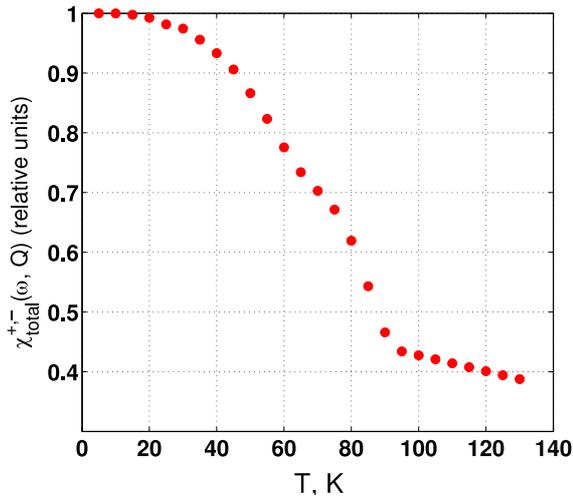}
\caption{Temperature dependence of the resonance peak intensity.}
\label{fig:TemperatureDependence}
\end{figure}

Finally, in Fig.\ref{fig:TemperatureDependence} we show the calculated
temperature dependence of the resonance peak using the temperature
dependence of the gap \newline
$\Delta(T) = \Delta_0 \tanh\left(1.76\sqrt{T_c/T - 1}\right)$ which was obtained previously\cite{mayer,eremin3}. 
The resonance peak intensity follows the temperature
dependence of the superconducting gap. Above $T_c$ the resonance mode continuously evolves
into paramagnon mode. We stress that the true resonance exists only below T$_c$ and the excitations above T$_c$ refers to the spectrum of the normal state. This result is in agreement with
the experimental observation found in  $YBa_{2}Cu_{3}O_{6.97}$ (see Fig.10 in
Ref. \cite{bourge} and also Fig.4(a) in Ref. \cite{bourge2}).

At the end of this section we shortly summarize the differences between our results 
and those obtained earlier either in a conventional RPA scheme or within a memory 
functions formalism (MFF). As it was pointed out in previous RPA studies, see Ref. \cite{reznik}, 
the conventional RPA scheme explains quite well the downward part of the dispersion of the 
spin excitation which originates in this case due to d-wave symmetry of the 
superconducting gap, but does not reproduce the upward one which should refer to the 
paramagnon dispersion. The origin of this is that the lifetime of the quasiparticles 
is not included within simple RPA. As a result, the paramagnons at higher energies are 
strongly damped due to a standard Landau damping mechanism. In our approach the Landau 
damping is suppressed by the kinematic "molecular" field which originates from local 
no-double occupancy constraint. The constraint reduces the phase space available for 
the particle-hole continuum to form and suppresses then the Landau damping. As a result 
the upper paramagnon part of the dispersion is more visible. 

At the same time the chains of equations of motion with a decoupling procedure employed 
by us and memory function formalism used by other groups  are closely related. In the 
memory function formalism one decouples the exchange part of the Hamiltonian in fashion 
similar to ours. The resulting dispersion of the localized excitations is then again 
referred to the  upward branch of the spin excitation seen in INS. To account for the 
superconductivity and presence of itinerant electrons, which give rise to the downward 
dispersion, one introduces the decoupling of the spin-excitation damping. This describes 
the decay of the spin excitation into an electron-hole pair. Here, superconductivity is 
introduced phenomenologically in the energy spectrum of the quasiparticles. One finds that 
the results for the spin excitations within memory function and equations of motion formalism are quite
similar, see Refs. \cite{sherman,prelovsek,vladimirov}. However, we believe that our approach dealing 
with the chains of equations of motion has certain advantages. In particular, it allows 
to reproduce wider range of experimental data on INS for $T<T_c$ (also including NMR and 
ARPES data) and, most importantly, we are able to describe both upward and downward dispersions 
(see Fig. 3) within a single analytical expression. Here, the origin of various terms have 
a transparent physics meaning and the parameters of the quasiparticle dispersion and the spin 
excitation spectrum are computed self-consistently.

\section{Concluding remarks}

 To conclude, in this paper we present the analysis of the spin
response in the superconducting cuprates taking into account both the local
and the itinerant component of the magnetic excitations coupled 
in a self-consistent manner. The numerical results show that the
obtained expression for the spin susceptibility, Eq. (\ref{eq:chiTotal}),
reproduces well the characteristic features of the experimental data in 
$YBa_{2}Cu_{3}O_{6+y}$ compounds near the optimal doping including the
dispersion of the spin excitations in the normal and superconducting state
as well its frequency and temperature dependence. While the structure of the
spin excitations in the normal state can be attributed to the 
overdamped localised magnetic modes, the strong feedback of the $d$-wave
superconductivity on the itinerant electrons reveals the formation of the
spin exciton with the characteristic downward dispersion of the spin
excitations in the superconducting state. In addition, the high energy spin
excitations still originates from the localised excitations. Remarkable that
both modes merge at the $\omega_{res}$ which is a result of the single pole
structure at \textbf{Q} in Eq. (\ref{eq:chiTotal}). Our analytical
results suggests the dual character of the spin response even in the optimally doped 
cuprates. \newline

The authors acknowledge A. Barabanov, M. Mali, P.F. Meier, A. Mikheenkov, N. Plakida, J. Roos, 
for useful remarks and discussions. This work was supported
in part by the Russian Foundation for Basic Research,
Grant 09-02-00777-a and Swiss National Science Foundation,
Grant IZ73Z0 128242.

\end{document}